\documentclass[10pt, a4paper, twocolumn, copyright, goog]{google}

\usepackage{listings,xcolor}
\usepackage{color,amsmath,enumitem}
\usepackage{language_formatting}
\usepackage{graphicx}
\usepackage[authoryear, sort&compress, round]{natbib}
\usepackage{xurl}
\keywords{Trusted Execution Environments, Federated Learning, Differential Privacy, Scalable Systems}

\title{Toward provably private learning from federated data}

\correspondingauthor{dalyk@google.com}

\author[1]{Katharine Daly}
\author[1]{Yu Xiao}
\author[1]{Zachary Garrett}
\author[1]{Brett McLarnon}
\author[1]{Jianpeng Hou}
\author[1]{Arun Ganesh}
\author[1]{Yanxiang Zhang}
\author[1]{Noriyuki Takahashi}
\author[1]{Haicheng Sun}
\author[1]{Yuanbo Zhang}
\author[1]{Timon Van Overveldt}
\author[1]{Daniel Ramage}

\affil[1]{\thepa{}{}}

\begin{abstract}
Federated Learning (FL) allows devices with private data to collaborate in training a shared model. We present a next-generation FL system based on Trusted Execution Environments (TEEs) that addresses operational challenges associated with earlier systems and provides externally verifiable central Differential Privacy (DP) guarantees for the first time while offering a better privacy-utility tradeoff. In our system, devices upload data encrypted with keys managed by a TEE-hosted Key Management Service (KMS). The uploaded data is cryptographically tied to a policy limiting the set of Python programs that may later process the data in server-side TEEs. External parties may inspect public transparency logs to observe the set of workloads allowed by these policies. Our experimental results show that the new system improves device coverage and favorably shifts privacy-utility curves by enabling collected data to be integrated into the server-side workload at a schedule that optimizes DP guarantees and is unaffected by device availability. Our new system has been productionized, enabling models for the Android Keyboard (Gboard) to be trained faster and achieve better accuracy under smaller, now externally verifiable privacy budgets in comparison to models trained using the prior system.
\end{abstract}

\begin{document}

\maketitle

\section{Introduction}
\label{introduction}

Federated Learning (FL) was introduced in 2016 \citep{mcmahan2017communication} as a machine learning technique for training models in a privacy-preserving, distributed manner across decentralized data. At Google, this original FL technique has been used to train models deployed on Android for applications including next-word prediction \citep{hard2018federated}, emoji suggestion \citep{ramaswamy2019federated}, smart text selection \citep{hartmann2021predictingtextselections}, and smart reply \citep{google2020messagesimprovessuggestions}. Apple \citep{paulik2021federated} also built an FL system for consumer devices that led to models used by Siri for personalized speech recognition and by the QuickType keyboard. Other groups have built cross-silo FL systems to train on medical records owned by various hospitals \citep{melba:2025:033:linardos} and to train fraud detection systems on financial data held by various banking partners \citep{roth2026privacy}.

In 2024, we introduced an evolved definition of FL \citep{daly2024federated} that focuses on privacy principles rather than a specific division of work between components:

\begin{quote}
\emph{\textbf{Federated learning} (FL) is a machine learning setting where multiple entities (clients) collaborate in solving a machine learning problem, under the coordination of a service provider. A complete FL system should enable clients to maintain full control over their data, the set of workloads allowed to access
their data, and the anonymization properties of those workloads. FL systems should provide appropriate
transparency and control to the users whose data is managed by FL clients.}
\end{quote}

Alongside this new definition, we also proposed ways in which future FL systems could raise the bar on each of four privacy principles (data minimization; data anonymization; transparency and control; and verifiability and auditability) compared to earlier FL systems. These four privacy principles were defined earlier in \citep{bonawitz2021federated}. In this new paper, we describe the FL system we built using Trusted Execution Environments (TEEs) enabled by AMD SEV-SNP \citep{amdsevsnp} and Intel TDX \citep{intelTDX} to achieve this higher privacy bar.

In addition to furthering FL system privacy properties, our new FL system also aims to address other practical limitations of earlier FL systems involving consumer devices. Earlier systems required computing model gradients on-device, which posed limitations on the size of models that could be trained. Various synchronization points across fleets of devices with unpredictable availability and heterogeneous compute resources also posed operational challenges \citep{kairouz2021advances}. Our new system allows devices to participate independently of each other and moves most computation server-side.

The remainder of this paper is organized as follows. Section \ref{related_work} discusses related work on TEEs. Section \ref{system_description} describes our new TEE-based FL system design. Section \ref{theoretical_comparison} compares our new and old systems with respect to various privacy considerations. Section \ref{experimental_results} includes experimental results comparing our new and old systems for various models. Section \ref{future} describes future work, and we conclude the paper in Section \ref{conclusion}.

\section{Related Work}
\label{related_work}
Trusted Execution Environments (TEEs) have increasingly been deployed on the server in recent years to support on-device features that require privacy but are computationally expensive. TEEs provide remote attestation capabilities as well as confidentiality of in-memory state and integrity of executed code, which together allow computations that might have previously been executed on-device to be moved server-side while maintaining privacy guarantees. Apple Private Cloud Compute \citep{apple2024privatecloudcomputesecurityguide}, Google Private AI Compute \citep{google2025googleprivateaicompute}, and Meta Private Processing \citep{meta2025privateprocessing} are all recent systems that run server-side LLM inference in TEEs or similar confidential computing environments to support a variety of mobile use cases requiring computationally expensive text summarization, multi-step reasoning, image generation, and alike. The specific privacy guarantees, security guarantees, and level of external inspectability vary across the systems. Current-generation TEEs are known to have limitations that affect their security guarantees \citep{Zhang2026}.

TEEs have also been used before in federated learning systems. Meta’s Papaya FL system \citep{huba2022papaya} is an asynchronous FL system that computes model gradients on-device and uses Intel SGX TEEs server-side to aggregate device updates. In contrast to their prior synchronous FL system, devices independently compute gradients and upload results, which are incorporated into the global model in a weighted manner in order to maintain high device utilization and avoid synchronization bottlenecks. Other FL systems have also used TEEs at the edge in addition to using them server-side. Intel’s OpenFL system \citep{foley2022openfl} is designed to run using Intel SGX enclaves at the edge to protect model IP and to prevent model poisoning by allowing the server-side TEE to validate the code that was run within the edge TEEs before incorporating updates.

We previously described a multi-purpose architecture for running externally verifiable workloads in server-side TEEs \citep{eichner2024confidential}. This work introduced several central, workload-agnostic components that allow devices to independently upload data to an intermediate location with the guarantee that the data may only be processed in the future by a set of server-side TEE workloads known at upload time. In \citep{cheu2025toward}, we described a production deployment of this system for workloads involving LLM structured summarization and DP aggregation, in which an untrusted coordinator orchestrates a data processing pipeline consisting of one or more server-side TEEs. In this paper, we describe a production deployment of the same system for workloads described using arbitrary Python code, which can include ML training and evaluation workloads. In this new deployment, which reuses the same central components, a server-side root TEE orchestrates execution of a python script across a collection of server-side worker TEEs.

\section{System Description}
\label{system_description}
\begin{figure*}[t]
    \centering
    \includegraphics[width=\textwidth]{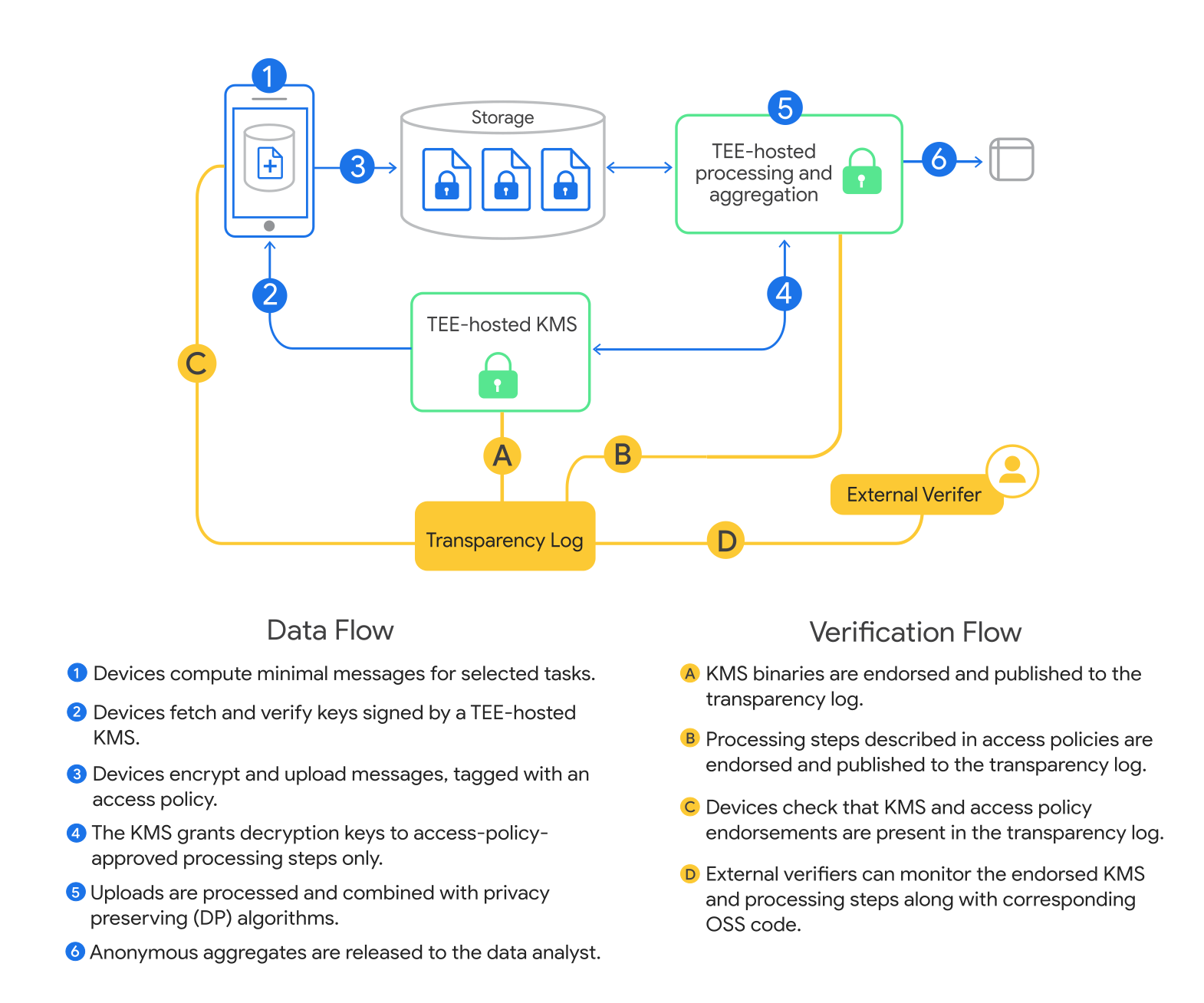}
    \caption{The data and verification flow in our TEE-based system for execution of arbitrary Python workloads.}
    \label{fig:system_flow}
\end{figure*}

Building on \citep{eichner2024confidential}, our system consists of an initial data upload stage followed by a server-side data processing stage. A \textit{workload author} specifies the custom logic that runs in each of these stages for a given use case, and a \textit{pipeline operator} triggers execution of the logic on the system. In many cases, the workload author and pipeline operator are the same entity but are distinct from the owner / operator of the system components. The two stages of the system are intermediated by a Key Management System (KMS) component that allows a chain of trust to be built from the client devices to the future server-side data processing stage before devices even upload data. Both the KMS and the server-side data processing stage utilize collections of TEEs. In our deployment of this system for workloads expressed using Python, which is depicted in Figure \ref{fig:system_flow}, the combination of workload-agnostic components (e.g. the KMS) as well as new data processing components allows new properties to be achieved:
\begin{itemize}
\item \textbf{Verifiable execution of arbitrary code within system-provided data processing TEEs.} Workload authors use a simple program API, which we describe in Section \ref{program_api}, to express the logic they want to execute in new server-side, system-provided data processing TEEs over sensitive device data. Devices upload encrypted data with the assurance that the uploaded data will only be able to be decrypted and processed at a later time by those specific programs. The programs from workloads authors and the system-provided runtime binaries that are used to execute them inside the TEEs are inspectable by external parties.
\item \textbf{Efficient, distributed execution.} The programs are executed in a `root' node in a graph of TEE-hosted compute nodes. The root node can delegate parallelizable units of work to worker nodes, verifying their integrity transitively.
\item \textbf{Proprietary per-user data processing.} If a workload author wants to keep a per-user data transformation private, it can be applied independently to each user's data without affecting the externally verifiable privacy guarantees, subject to known limitations such as side-channel observability.
\item \textbf{Dynamic parameter loading.} Programs can rely on parameters or other pieces of data that are dynamically configured at runtime without affecting privacy guarantees.
\item \textbf{Privacy-neutral, automated failure recovery.} Mechanisms are available to let programs recover from failures without losing already-completed work and without impacting privacy guarantees.
\end{itemize}

\subsection{Program API}
\label{program_api}
Workload authors express the logic they want to run during the server-side data processing stage by writing a Python program that contains the following signature:

\begin{minipage}{\linewidth}
\begin{lstlisting}[language=Python]
def trusted_program(external_handle: ExternalHandle):
    # custom workload logic goes here
\end{lstlisting}
\end{minipage}

When the server-side data processing stage runs, a TEE binary (the `root' compute node) hosting a Python runtime calls the \texttt{trusted\_program} function with an initialized \texttt{ExternalHandle} instance. The \texttt{ExternalHandle} instance provides runtime I/O capabilities to the program:
\begin{itemize}
\item \textbf{Access to uploaded data.} The program can use \texttt{ExternalHandle} to access the uploaded data, which it is responsible for parsing.
\item \textbf{Sideloading of information.} The program can use \texttt{ExternalHandle} to load arbitrary information at runtime. This functionality is useful both for runtime parameterization of programs as well as for incorporating per-user proprietary mapping logic into an otherwise open source program. The program is responsible for interpreting the sideloaded information, which is provided by the pipeline operator at runtime. While client data and sideloaded information are both runtime inputs to the program, they differ in that client data is only decryptable inside the TEE while sideloaded information is not encrypted.
\item \textbf{Controlled release of results.} The program can use \texttt{ExternalHandle} to emit plaintext results from the TEE to the pipeline operator for analysis. It is the responsibility of the workload author to ensure that any result that is released has been appropriately anonymized to meet the workload’s privacy claims. As will be discussed more in Section \ref{chain_of_trust}, external verifiers can view open source programs to validate that this is indeed the case.
\item \textbf{Saving and loading of recovery information.} 
The program can use \texttt{ExternalHandle} to periodically emit information from which the program can be restarted at a later time. The format of this information is controlled by the program. The information is signed and KMS-encrypted before leaving the TEE, ensuring that only future TEEs running the same program and binary code can read and recover from it. Replay attacks by server-side pipeline operators attempting to restore old recovery information are prevented via the mechanism we describe in Section \ref{pipeline_state}.
\end{itemize}

See Listing \ref{lst:basic_program} in the Appendix for an example of a program that demonstrates using \texttt{ExternalHandle} to achieve all these types of I/O.

\subsection{Distributed Execution}
\label{distributed_execution}
In the simplest deployment of our system for execution of arbitrary Python workloads, the server-side data processing stage consists of a singular TEE that executes the \texttt{trusted\_program} function. For compute-heavy workloads, however, a root TEE executes the \texttt{trusted\_program} function and delegates portions of the work to a collection of worker TEEs.

In cases where distributed execution is desired, the program itself must provide hints for how work should be delegated by the root TEE to worker TEEs. For this, we utilize Federated Language \citep{google2024federatedlanguage}, a framework-agnostic orchestration language and compiler infrastructure for expressing distributed computations. The portions of programs that contain Federated Language intrinsics (e.g. \texttt{federated\_broadcast}, \texttt{federated\_map}, \texttt{federated\_mean}, etc) may be distributed by the root TEE to worker TEEs for parallelized execution, whereas the portions of programs that do not use Federated Language are executed entirely by the root TEE.

\subsection{Verifiability and Transparency}
\label{verifiability_and_transparency}
Before devices upload data, they know the full set of server-side workloads that may run over the uploaded data in the future. This is made possible by a chain of trust brokered by the KMS that extends from the device through the KMS to future root and worker TEEs. As described in more detail in \citep{cheu2025toward}, the KMS consists of a cluster of TEEs implementing the RAFT consensus protocol. The KMS maintains a long-lived, rollback-protected key-value database that stores encryption/decryption key pairs and pipeline-specific state. In this section, we first describe how the chain of trust is established and then detail the pipeline state we store in order to provide privacy guarantees for execution of arbitrary Python programs.

\subsubsection{Chain of Trust}
\label{chain_of_trust}
As described in \citep{cheu2025toward}, during the upload stage devices establish trust in a KMS-provided public encryption key by (1) validating that the KMS's TEE software has been recorded in a public transparency log, and (2) validating that the \textit{access policy} that describes how the data will be allowed to be processed has also been published to that transparency log. The device then encrypts and uploads its data to an intermediate storage location using the public key and cryptographically ties it to the access policy. The KMS will only provide decryption keys to TEE binaries allowed in the access policy.

As shown in Listing \ref{lst:access_policy} in the Appendix, for workloads that execute an arbitrary Python program, the access policy directly embeds the Python program with the \texttt{trusted\_program} function as well as the binary hashes of the root and worker binaries described in Section \ref{distributed_execution}. Hence, only that specific Python program, executed by those specific root and worker binaries, can access the uploaded data inside the TEEs. The Python program becomes open source when the access policy is published to a public transparency log, along with the software for the root and worker binaries. External auditors can view how the program establishes DP guarantees. While parameters or per-user data processing logic delivered at runtime via the sideloading capability of \texttt{ExternalHandle} do not become open source, auditors can inspect how these are used to conclude that DP guarantees are maintained.

The last link in the chain of trust extends from the root TEE to the worker TEEs. Prior to delegation of any work, root and worker TEEs establish a bidirectional, end-to-end encrypted channel using an implementation of the Noise Protocol \citep{perrin2018noiseprotocol} running over an untrusted gRPC channel. The root TEE validates the worker TEE's attestation measurements during the establishment of the Noise channel.

\subsubsection{Pipeline State}
\label{pipeline_state}
In addition to managing encryption/decryption keys to ensure data is processed by the correct server-side pipelines, the KMS also holds pipeline state. Upon execution of a pipeline, the KMS provides the latest pipeline state for that pipeline to the data processing TEEs, which use the pipeline state to store relevant information across executions. Unencrypted results can only be released from the root TEE by atomically updating the corresponding pipeline state at the same time.

To allow programs to recover from transient failures during execution without requiring extra privacy budget, the pipeline state that the root TEE stores in the KMS for a given Python program identifies the encrypted recovery information that was written atomically alongside the most recently released unencrypted result. Suppose a program is written to execute 1000 rounds of training over batches of user data, releasing an anonymized model checkpoint every 10 rounds and storing new recovery information at the end of every round. Recovery information released via \texttt{ExternalHandle} is encrypted with a key bound to the access policy by the KMS and also signed by the root TEE, and for a typical training program it would include the latest checkpoint, latest round number, and the round participation schedule. If a root TEE executing this training program fails in the middle of round 206, a new instance of the root TEE will permit the program to begin executing from the recovery information stored at the end of rounds 200 through 205. The new root TEE will not permit recovery from earlier rounds because doing so would cause unencrypted data to be re-released for earlier rounds, which for non-deterministic programs can have unintended privacy impact (e.g. if the released data for a round is a function of random noise, then being able to inspect multiple released results for the same round would produce more information about the uploaded data than intended).

\section{System Privacy Comparison}
\label{theoretical_comparison}
In this section, we compare privacy considerations of running FL training workloads on our prior FL system vs our new TEE-based system. Some aspects of the new design lead directly to improvements in the privacy principles we introduced in Section \ref{introduction}, whereas other aspects of the new design produce new headroom that can be allocated towards either stricter differential privacy (DP) guarantees or improved utility.

\subsection{Data Minimization and Anonymization}
In our old system, the data uploaded to the server for FL training workloads was model gradients that were computed on-device. Because the server was untrusted and could directly access client uploads, techniques such as Secure Aggregation (SecAgg, \citealt{bonawitz2017practical}) and its successor Willow \citep{bell2025willow} coupled with distributed DP \citep{kairouz2021distributed} were deployed to limit the information the server could observe from client uploads.

In our new system, the data uploaded to the server for FL training workloads is encrypted training data. Uploading raw training data rather than client gradients allows the DP training loop to be executed exclusively server-side after a one-time client upload step, simplifying the execution of the workload. The pipeline operator is unable to access the raw training data. Instead, the pipeline operator is only able to view results generated by workloads represented in the open source access policy which is cryptographically tied to the client uploads. This assertion holds even when the pipeline operator is untrusted due to the chain of trust described in Section \ref{chain_of_trust} that links the devices to the server-side processing TEEs. The uploaded data is also protected by a TTL that is enforced on a best-effort basis by the KMS, meaning the uploaded data cannot be decrypted by any TEE, including a TEE whose binary hash is represented in the associated access policy, after a certain timeframe.

\subsection{Transparency, Control, Verifiability, and Auditability}
In our old system, device owners could opt out of participating in FL training workloads \citep{googlelearnhowgboardgetsbetter}, and with substantial effort they could inspect the on-device binaries to check whether device-level privacy protections such as SecAgg were being applied as claimed. Device owners were unable to verify the logic being run server-side.

In our new system, device owners can still opt out of participating in FL training workloads, and their visibility into what logic is running as part of the server-side workload is drastically improved. Prior to uploading, devices are able to know the full set of server-side workloads that may run in the future over uploaded data. Because the access policies describing server-side workloads are uploaded to transparency logs, third-party auditors can also know the full set of potentially active binaries and workloads even without having a device that participates in the training process. The binary hashes included in the access policies describe reproducibly buildable TEE containers that are built from OSS source code.

\subsection{Privacy-Utility Curves}
\label{theoretical_privacy_utility}
Using our old system, we trained models supporting various Android keyboard (Gboard) features using DP-FTRL with clients permitted to contribute only once within a given time period (e.g. once per 144hr) \citep{xu2023federated}. This time interval had to be carefully tuned to account for diurnal changes in device availability and desired cohort sizes. Given the observed round completion timestamps and the time period that controlled contributions, we were able to calculate the minimum number of rounds separating participations from an individual device (\emph{minSep}) and the maximum number of rounds in which an individual device could have participated (\emph{maxP}), which were then used alongside other DP-FTRL settings to calculate the \emph{z}CDP for the training process.

In our new system, we collect all device uploads prior to executing a FL training workload on the server, which allows us to optimize MF-DP-FTRL \citep {choquette2023amplified} parameters and client participation patterns at runtime and favorably shift the privacy-utility curve. We are able to have devices participate an equal number of times and can also achieve the maximum possible \emph{minSep} based on the number of available uploads and the target number of rounds. From this maximum possible \emph{minSep} value, we derive a BLT with improved utility \citep{mcmahan2024hassle} and then determine the noise multiplier based on the target \emph{z}CDP (or vice-versa). Section \ref{experimental_privacy_utility} includes quantitative results showing the extent to which we were able to move the privacy-utility curve for a specific model.

\section{Experimental Results}
\label{experimental_results}
\begin{figure*}[t]
    \centering
    \includegraphics[width=0.95\textwidth]{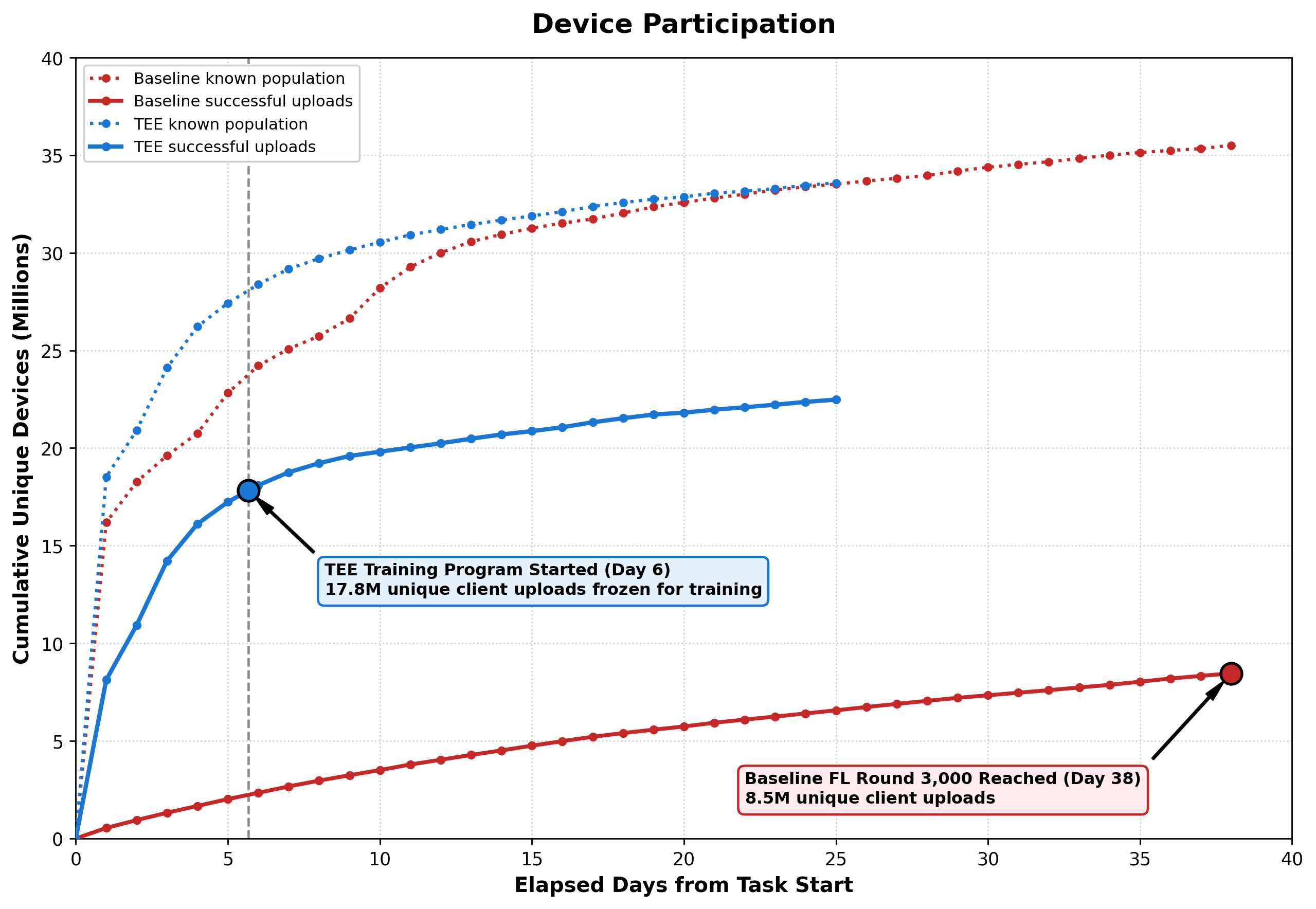}
    \caption{The TEE-based system is able to incorporate more unique devices into training than the old baseline system. The dotted curves show the cumulative unique devices observed by the server over time since beginning training (or in the TEE case, beginning to collect uploads) for a Japanese language model. In 3000 rounds of training over 38 days, the baseline case incorporated 8.5M unique devices into training. In approximately 6 days of data collection, the TEE case collected 17.8M uploads, all of which were used in the training process that was subsequently launched.}
    \label{fig:device_coverage}
\end{figure*}

\begin{figure*}[t]
    \centering
    \includegraphics[width=0.8\textwidth]{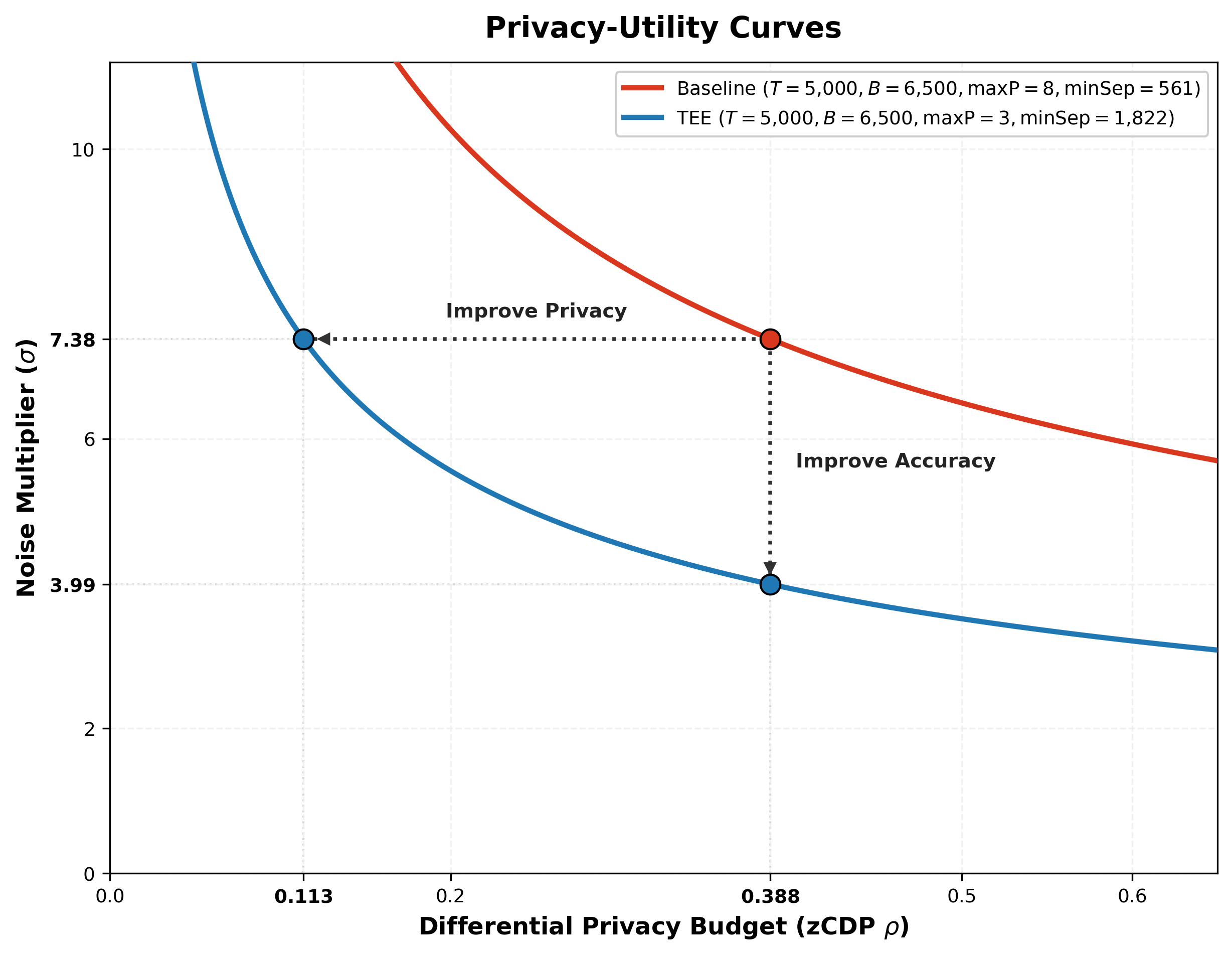}
    \caption{The TEE-based system provides additional headroom that can be allocated to a smaller privacy budget, a reduced noise multiplier, or both. Both curves depict the possible \emph{z}CDP and noise multiplier combinations possible at T=5000 rounds for an English language model trained on both the new TEE-based system and the baseline system with cohort size B=6500. The \emph{maxP} (maximum number of rounds in which a single device participated) and \emph{minSep} (minimum number of rounds separating participations by a single device) impact the sensitivity and thus the consumed privacy budget.}
    \label{fig:privacy_utility}
\end{figure*}

We have used the new TEE-based system to train Gboard models over various populations. In this section we present experimental results showing key differences between training models in the new TEE-based system vs the older baseline system.

\subsection{Improved Device Coverage}
\label{device_coverage}
The new TEE-based system is able to train over a greater number of unique devices. Figure \ref{fig:device_coverage} displays this improvement for a Japanese language model. For both the baseline and TEE systems, the number of cumulative unique \emph{available} devices (devices that are connected to WiFi, are plugged in, and have sufficient battery) that were observed over time by our servers was similar. However, the TEE system was able to incorporate contributions from a larger fraction of available devices.

In the baseline case, throughout 3000 rounds of training over 38 days with a cohort size of 6500, only 8.5M of 35.5M available devices actually contributed data to the training process (23.9\%). In the baseline system, devices that are available more often or are available at unpopular times are more likely to be selected for participation by the server, which forms cohorts of devices to participate in a round of training. Of the devices that were available at some point during the training process but did not contribute, 20.5M never received a workload to run from the server and 6.5M received a workload at least once but were never able to complete it successfully due to interruption, such as availability conditions becoming unmet.

In the TEE case, the server-side training stage was started after receiving 17.8M uploads over approximately six days (additional uploads received after the training stage begins cannot be incorporated, although we did continue to run the upload task in this case to observe its behavior). Due to the TTL on the decryption keys, users of the system must balance between waiting for more devices to upload vs giving the training process sufficient time to run. For example, if a server-side program begins on day $y$ and runs over data collected between days $x$ to $y$ that was encrypted with KMS keys with a 30 day TTL, then the program is only permitted to produce outputs for $30-(y-x)$ days. The model in the TEE case was configured to train with MF-DP-FTRL for 3000 rounds with a cohort size of 6500. Because the TEE training program has on-demand access to all previously collected uploads, it is able to optimize their participation pattern by making the \emph{minSep} as large as possible, as described in Section \ref{theoretical_privacy_utility}. This also has the side effect that all 17.8M uploads were incorporated into training and participation was unbiased across these 17.8M uploads. Of the devices that were available but did not upload before the server-side training stage was started, the vast majority simply did not have eligible data. Unlike in the baseline case, devices rarely failed to upload to due interruption. This is unsurprising given that the amount of work devices need to do to prepare their upload is significantly reduced in the TEE case.

\subsection{Shifted Privacy-Utility Curves}
\label{experimental_privacy_utility}
The TEE-based system shifts privacy-utility curves inward, allowing greater privacy to be achieved for the same utility, or vice-versa. Figure \ref{fig:privacy_utility} displays the headroom gains achieved by an English language model. In both the baseline and TEE cases, the model was trained using MF-DP-FTRL with a cohort size of 6500.

In the baseline case, the noise multiplier was fixed at 7.38, and the model trained for over 8616 rounds over 85 days. Devices were eligible to participate every 144 hrs (6 days). For a given checkpoint, the \emph{maxP} and \emph{minSep} were derived based on this 144 hr window and the timestamps of the earlier checkpoints. The consumed privacy budget (\emph{z}CDP) was then calculated based on the \emph{maxP}, \emph{minSep}, noise multiplier, and a preset BLT matrix.

In the TEE case, we targeted a \emph{z}CDP of 0.232. The noise multiplier required to meet this max \emph{z}CDP at 5000 rounds was determined at runtime to be 5.16 based on the number of collected uploads (11.8M), which was also used to compute the optimal BLT matrix. A \emph{z}CDP of 0.232 would have required a significantly higher noise multiplier (9.54) in the baseline case at the same number of rounds.

Comparing the \emph{maxP} and \emph{minSep} at 5000 rounds in the baseline case (\emph{maxP}=8, \emph{minSep}=561) to the TEE case (\emph{maxP}=3, \emph{minSep}=1822) shows the source of the additional headroom. The function mapping the noise multiplier to utility depends on the BLT, and as the BLT is optimized for lower \emph{maxP} and higher \emph{minSep}, the mapping function becomes more favorable, meaning Figure \ref{fig:privacy_utility} actually undersells the improvement in utility. The optimized participation patterns that were achieved in the TEE case were made possible by the two-stage design, which facilitates on-demand access to a set of device uploads.

\subsection{Live A/B Testing}
We conducted an A/B experiment on a subset of Gboard's production population to evaluate checkpoints from an English language model trained using the new TEE-based system. Gboard's typing decoder, which powers auto-correction, word completion, and next-word prediction features, uses a language model in two separate stages: Neural Search Space (NSS) for first-pass candidate generation and On-The-Fly (OTF) rescoring for second-pass candidate re-ranking \citep{zhang2024neural}. Each arm of the experiment included 3.5M devices. Large gains were made on privacy, with the best TEE-based arm requiring a 3x smaller privacy budget ($z$CDP=0.215) in comparison to the production English language model ($z$CDP=0.641 when adjusted for the same MF-DP-FTRL mechanism) while maintaining neutral performance on key usability metrics such as Words Per Minute (capturing users' typing speed) and Words Modified Ratio (capturing whether users modify Gboard's suggestions). The model in the TEE-based case was trained in 3 weeks, compared to 2 months of training for the previous production model.

\section{Future Work}
\label{future}
\begin{figure*}[t]
    \centering
    \includegraphics[width=\textwidth]{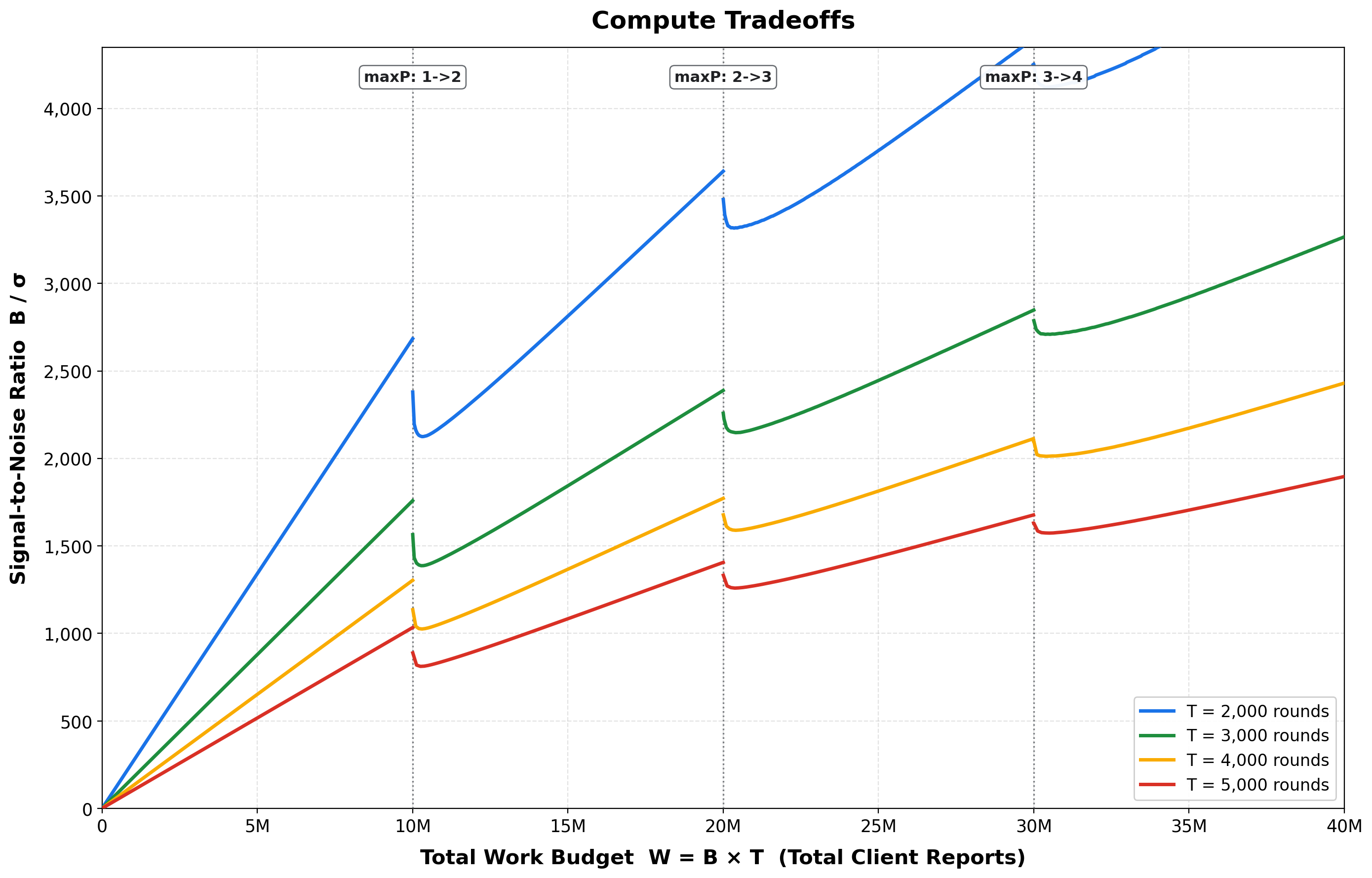}
    \caption{The signal-to-noise ratio (cohort size, $B$, divided by noise multiplier, $\sigma$) improves for a constant work budget (represented by the number of rounds, $T$, multiplied by the cohort size, $B$) as the cohort size increases. However, at the transition points where an increased work budget causes the maximum participations (\emph{maxP}) of a given device to increment, the signal-to-noise ratio jumps down as more noise is suddenly needed to address the additional possible participation. This plot was computed for a population size of 10M and a target \emph{z}CDP of 0.5, and the BLT matrices were optimized using a buffer size of 3.}
    \label{fig:compute_tradeoffs}
\end{figure*}

In this section, we explore possible future work spanning both infrastructure, operational, and algorithmic improvements.

\subsection{Infrastructure Scaling}
The new TEE-based system already relocates some previously device-side computation to the server, where compute availability is usually less limited. For training workloads, work can be distributed from the root TEE to the worker TEEs, allowing computation parallelism which improves round times significantly when training on as few as 14 machines as in the experiments above.

So far we have used the new TEE-based system to train models up to 10M parameters in size. To support training much larger models efficiently, worker TEEs may need to utilize GPUs, and additional communication bottlenecks may need to be resolved between the root and workers.

\subsection{Hyperparameter Optimization}
As illustrated in Figure \ref{fig:compute_tradeoffs}, choosing the number of rounds and cohort size optimally given a population size and privacy budget is a model-specific optimization problem. A sudden drop in the signal-to-noise ratio (the cohort size, $B$, divided by the noise multiplier, $\sigma$) occurs when the number of maximum participations (\emph{maxP}) for a given device increments, since more noise is needed to offset the potential additional participation. It is thus suggested to target a work budget (approximated by $T * B$, where $T$ is the number of rounds and $B$ is the batch size) just prior to such a \emph{maxP} transition. At a constant work budget, increasing the cohort size improves the signal-to-noise ratio. However, increasing the cohort size also has diminishing returns on the per-step convergence rate \citep{charles2021large}, meaning there is a limit to how much the cohort size should be increased within a fixed compute budget, given the primary goal of convergence. 

Additional experiments are needed to find the optimal hyperparameter configurations given these tradeoffs for the Gboard models. For the experiment results presented in Section \ref{experimental_results}, only one configuration was tested per model. Recent scaling laws for DP model training that balance privacy, compute, and population constraints may provide useful guidance \citep{mckenna2025scaling}. Note that the server-side Python program itself could explore different hyperparameter configurations, implementing early stopping and dynamic selection of the next candidate, rather than requiring the pipeline operator to run multiple server-side workloads in parallel.

\subsection{Additional Privacy-Utility Headroom}

Currently the pipeline operator is able to observe which uploads (but not their contents, since the uploads are encrypted) are being used for each round of training. If this information was obfuscated from the pipeline operator, we could benefit from privacy amplification by sampling, improving the externally verifiable privacy guarantees for models trained with MF-DP-FTRL. In other words, the provable consumed privacy budget would decrease while the noise multiplier remains constant. There are at least two ways in which information about uploads could be obfuscated from the pipeline operator: data could be verifiably shuffled, or the server-side workload could over-select the data it needs for each round (i.e. the pipeline operator would know that a subset of a given set of uploads was used for a round, but it wouldn't know the exact subset the program chose to use).

\subsection{Additional Use Cases}
The TEE-based system described in Section \ref{system_description} supports execution of arbitrary Python programs. Our first use case was model training, but the same infrastructure can be used for additional use cases such as generation of synthetic data, inference, etc. We have begun exploring other such use cases and also hope to be able to run pipelines that use both the data processing TEE containers described in this paper and those described in our earlier work \citep{cheu2025toward}.

\section{Conclusion}
\label{conclusion}
We have described our TEE-based FL system, which provides externally verifiable Differential Privacy (DP) guarantees while also improving other aspects of FL training in comparison to earlier systems. The two-stage approach of collecting uploads prior to running server-side training allows private device data to be incorporated into the training process at a schedule optimized for DP algorithms, resulting in more favorable privacy-utility curves. In an live A/B test for the Google Keyboard (Gboard) typing decoder, which provides auto-correction, word completion, and next-word prediction features, a TEE-trained English language model that was trained 3x faster and that used a 3x smaller privacy budget than the current production model maintained parity on key usability metrics. Future work includes further optimizing training hyperparameters, uncovering even more privacy-utility headroom, and scaling to larger models.

\section{Acknowledgments}
The authors thank the following additional contributors to this effort: Zachary Charles, Stefan Dierauf, Nova Fallen, Xiaojuan Fang, Emily Glanz, Suxin Guo, Liyang Jiang, Yingjie Liu, Wenzhi Mao, Brendan McMahan, Michael Reneer, Maya Spivak, Heng Su, Yun Wang, and Chunxiang (Jake) Zheng.

The authors also thank Brendan McMahan for feedback on this paper.

\bibliography{references.bib}

\onecolumn

\appendix
\section{Supplemental Listings}
\begin{figure}[h!]
\begin{lstlisting}[caption={An example program that utilizes the Program API. During each round of execution, the program selects from the available data, parses the selected data using a sideloaded proto definition, adds the extracted values to a cumulative sum, releases a noised version of the cumulative sum, and stores encrypted recovery information.}, label={lst:basic_program}]
from fcp.confidentialcompute.python import external_service_handle

# Entrypoint for TEE-hosted program.
def trusted_program(
    external_handle: external_service_handle.ExternalServiceHandle,
) -> None:

  # Define constants.
  ...

  # Helper function for parsing data.
  def parse_tensor(...):
    ...

  # Sideload the proto definition that describes the upload format.
  fds = descriptor_pb2.FileDescriptorSet()
  with open(external_handle.get_filename_for_config_id(MY_PROTO_ID)) as f:
    fds.ParseFromString(f.read())

  # If recovery information is available (which may be the case if the program
  # was previously interrupted), use it to continue from where the program left
  # off.
  recovery_val = external_handle.restore_recovery_info(RECOVERY_ID)
  if recovery_val is not None:
    starting_round_index, cumulative_sum = tuple(recovery_val.decode())
    starting_round_index += 1
  else:
    starting_round_index = 0
    cumulative_sum = 0

  for round_index in range(starting_round_index, num_rounds):
    # Select available clients at random for this round.
    selected_ids = random.sample(external_handle.blob_ids, NUM_CLIENTS_PER_ROUND)

    # Fetch and parse the data from the selected clients. Add it to the
    # cumulative sum.
    for id in selected_ids:
      tensor = external_handle.resolve_blob_id_to_tensor(id, KEYNAME)
      for data in parse_tensor(tensor, fds, PROTO_MESSAGE_NAME):
        cumulative_sum += data[MY_FIELD]

    # Release a noised sum for this round while also saving the round index and
    # cumulative sum as encrypted intermediate recovery information.
    noised_sum = cumulative_sum + random.randint(-10, 10)
    new_recovery_val = bytes((round_index, cumulative_sum))
    external_handle.save_recovery_info(new_recovery_val, RECOVERY_ID, [(noised_sum, f"round_{round_index}")])
\end{lstlisting}
\end{figure}

\begin{figure}[t]
\begin{lstlisting}[caption={The access policy that is cryptographically attached to uploaded data contains the inlined Python program as well as the binary hashes of the containers that are allowed to process the data.}, label={lst:access_policy}]
pipelines {
  # The KMS tracks pipeline state by this key. At execution time, the root container
  # receives the current pipeline state associated with the key and validates the 
  # state to ensure that the program is either running for the first time or
  # continuing from the last registered intermediate recovery information.
  key: "program_name"
  value {
    instances {
      transforms {
        # The program may read device uploads (id 0) and intermediate recovery
        # information (id 1) and may output encrypted intermediate recovery 
        # information (id 1) and unencrypted anonymized results (id 2).
        src_node_ids: 0
        src_node_ids: 1
        dst_node_ids: 1
        dst_node_ids: 2
        
        # These reference values store the expected digests and measurements for
        # the root container at four layers of the stack:
        # -Root Layer: Stage0 firmware digests & AMD SEV-SNP min SVN thresholds
        # -Kernel Layer: Kernel/initramfs digests + regex-enforced boot cmdline
        # -System Layer: Cryptographic digests of the trusted base OS image
        # -Container Layer: SHA digests of the root container binary
        # The KMS will only provide decryption keys to a root container instance
        # that matches these reference values.
        application {
          reference_values { /* oak.attestation.v1.ReferenceValues */ }
        }

        # These configuration constraints are enforced by the root container at
        # runtime.
        config_constraints {
          type_url: "type.googleapis.com/fcp.confidentialcompute.ProgramExecutorTeeConfigConstraints"
          value {
            # The Python program that is allowed to run. The root container will
            # not allow any other program to be executed.
            program: """
              # The maximum allowed zCDP enforced by the program.
              _MAX_ZCDP = 0.58
              
              def trusted_program(external_handle):
                ...
                if (computed_zcdp > _MAX_ZCDP):
                  raise ValueError(...)
                ...
            """
            
            # These reference values store the expected digests and measurements
            # for the worker containers at four layers of the stack:
            # -Root Layer: Stage0 firmware digests & AMD SEV-SNP min SVN thresholds
            # -Kernel Layer: Kernel/initramfs digests + regex-enforced boot cmdline
            # -System Layer: Cryptographic digests of the trusted base OS image
            # -Container Layer: SHA digests of the worker container binary
            # The root container will not delegate work to any worker container
            # instance that does not match these reference values.
            worker_reference_values { /* oak.attestation.v1.ReferenceValues */ }
          }
        }
      }
    }
  }
}
\end{lstlisting}
\end{figure}

\end{document}